\documentclass[preprint,amsmath,amssymb,11pt,english,aps,showpacs,showkeys]{revtex4}
\usepackage{graphicx}
\usepackage{graphics}
\usepackage{amsmath}
\usepackage{dcolumn}
\usepackage{amssymb}
\usepackage{bm}
\usepackage[latin1]{inputenc}
\newcommand{\be}{\begin{equation}}
\newcommand{\ee}{\end{equation}}
\newcommand{\bea}{\begin{eqnarray}}
\newcommand{\eea}{\end{eqnarray}}

\begin{document}
\title{On the alternative formulation of the three-dimensional noncommutative superspace}
\author{F. S. Gama, J. R. Nascimento, A. Yu. Petrov}
\email{fgama, jroberto, petrov@fisica.ufpb.br}
\affiliation{Departamento de Física, Universidade Federal da Paraíba, Caixa Postal 5008, 58051-970, João Pessoa, PB, Brazil}

\begin{abstract}
In this paper we propose a new version for the noncommutative superspace in 3D. This version is shown to be convenient for performing quantum calculations. In the paper, we use the theory of the chiral superfield as a prototype for possible generalizations, calculating the one-loop two-point function of a chiral supefield and the one-loop low-energy effective action in this theory.
\end{abstract}
\maketitle
The superspace, with no doubts, is a key idea of the supersymmetry concept.  By its essence \cite{Superspace}, the superspace is an extension of the usual Minkowski spacetime involving additional fermionic coordinates. Within the traditional supersymmetry, these extra coordinates obey the Grassmannian anticommutation relations, and the superfields are introduced as functions on the superspace. Further, all well-developed methodology of quantum field theories can be directly generalized for the superfields which allowed to elaborate supersymmetric extensions for the scalar field theories, gauge theories and gravity, and to study the properties of these supersymmetric theories on a quantum level.

The important step in the generalization of the superspace methodology was carried out in \cite{FL} where the non(anti)commutativity for the fermionic coordinates was proposed.  Further, the concept of the noncommutative superspace was formulated in a closed form in \cite{KPT}. Following it, one should suggest that the fermionic superspace coordinates obey the Clifford anticommutation relations instead of the Grassmann ones.  It was showed in \cite{KPT} that this formulation can be developed only by paying a price of a partial supersymmetry breaking. As a consequence, the non-anticommutative deformation of ${\cal N}=1$ supersymmetric theory implies a theory with ${\cal N}=\frac{1}{2}$ supersymmetry \cite{Sei}. One can note that, from a formal viewpoint, introduction of the Clifford anticommutation relations, involving the constant symmetric matrix, implements some special form of the Lorentz symmetry breaking in supersymmetric field theories different from the Berger-Kostelecky approach \cite{BK} where the deformation of the supersymmetry algebra is introduced in other way. At the same time, the Clifford anticommutation relations naturally imply formulating the Moyal product for the superfield, which, however, unlike the usual Moyal product, contains only a finite number of terms. Among the important features of this approach, one should also mention that the formulation of the noncommutative superspace is, first, essentially based on chiral representation of the supersymmetry algebra, second, is consistent only in the Euclidean space \cite{KPT}.

Further, the formalism of ${\cal N}=\frac{1}{2}$ supersymmetry was successfully applied for formulation of superfield models on the noncommutative superspace \cite{Sei} and shown to possess some deep motivations from the superstring theory \cite{BerSe}. Afterwards, the perturbative methodology was successfully applied for different ${\cal N}=\frac{1}{2}$ supersymmetric theories such as the ${\cal N}=\frac{1}{2}$ Wess-Zumino model whose renormalization properties have been discussed in \cite{NCWZ,NCWZ1, Spurion}, ${\cal N}=\frac{1}{2}$ supersymmetric gauge theories whose formulation and renormalizability have been studied in \cite{Penati,Penati1} (see also \cite{Ito,Ito1,Ito2,Ito3} for discussions on this issue), and extended supersymmetric theories, in particular, those ones formulated in terms of the harmonic superspace \cite{FerrIv,FerrIv1}. The calculation of the superfield effective potential for this kind of theories \cite{BBP,BBP1,BBP2,BBP3} also deserves to be mentioned.

At the same time, the superspace formulation of the three-dimensional field theories is well defined \cite{Superspace}. Therefore, the generalization of the concept of the noncommutative superspace for the three-dimensional case seems to be very natural and interesting. However, the natural problem arising in this study is well known \cite{earlier}: while in the four-dimensional case one has two sets of supersymmetry generators $Q_{\alpha}$ and $\bar{Q}_{\dot{\alpha}}$, so, deformation of the anticommutation relation between the fermionic coordinates implies only a partial supersymmetry breaking, in the three-dimensional case such a deformation results in a complete supersymmetry breaking. It was suggested in \cite{earlier} by some of us that a natural way to circumvent this difficulty would consist in using of the  extended, ${\cal N}=2$ supersymmetric theories from the very beginning, with further this extended supersymmetry is partially broken. However, in \cite{earlier}, as well as further in \cite{Faizal,Faizal1,Faizal2} where some alternative ways to realize this idea have been proposed, only some studies at the classical level have been carried out. At the same time, the use of the formulation of ${\cal N}=2$ three-dimensional superspace proposed in \cite{Hitchin}, and further applied for the quantum calculations in \cite{BS,Buch1,BMS} and \cite{earlierHD}, seems to be much more advantageous for development of the noncommutative superspace approach, due to the similarity of the ${\cal N}=2$ supersymmetry algebra in three-dimensional theories within this formulation and the ${\cal N}=1$ supersymmetry algebra in the four-dimensional theories.  Therefore, in this paper we formulate the noncommutative superspace with the use of the superspace formulation proposed in \cite{Hitchin}.

We should note nevertheless that, while these supersymmetry algebras are similar in many aspects, with the main of them is the presence of two independent types of the spinors, and hence of two types of spinor derivatives, there are essential differences between the four- and three-dimenional situations. The main of them is that while in $D=4$ there exist two spinor representations of the Lorentz group (undotted and dotted one), in $D=3$ there is only one spinor representation, which implies, first of all, in the presence of the new identities for spinor derivatives which have no analogues in the four-dimensional case. This makes our consideration and all calculations to be essentially different from the four-dimensional superfield theories.

We start with the following three-dimensional SUSY algebra treated in \cite{earlierHD}:
\bea
\{Q_{\alpha},Q_{\beta}\}=\{\bar{Q}_{\alpha},\bar{Q}_{\beta}\}=0; \quad\, \{Q_{\alpha},\bar{Q}_{\beta}\}=i\partial_{\alpha\beta}\equiv P_{\alpha\beta}.
\eea
The first step consists in constructing the analogue of the chiral representation for the SUSY algebra different from that one used in \cite{earlierHD} (we note that the chiral representation plays a crucial role for formulation of the fermionic noncommutativity, see \cite{KPT}). Therefore, by using the chiral coordinates $z^M=(y^{\alpha\beta},\theta^\alpha,\bar\theta^\beta)$, where $y^{\alpha\beta}=x^{\alpha\beta}+\frac{i}{4}(\theta^\alpha\bar\theta^\beta+\theta^\beta\bar\theta^\alpha)$, the new SUSY generators consistent with this algebra are:
\bea
\label{Q's}
Q_{\alpha}=i\partial_{\alpha};\quad\, \bar{Q}_{\beta}=i(\bar{\partial}_{\alpha}-\theta^{\beta}i\partial_{\beta\alpha}).
\eea
Here and further, $\partial_{\alpha}\equiv\frac{\partial}{\partial\theta^{\alpha}}$, $\bar{\partial}_{\alpha}\equiv\frac{\partial}{\partial\bar{\theta}^{\alpha}}$, and $\partial_{\alpha\beta}\equiv\frac{\partial}{\partial y^{\alpha\beta}}$. The corresponding supercovariant derivatives are
\bea
\label{deriv}
D_{\alpha}=\partial_{\alpha}+\bar{\theta}^{\beta}i\partial_{\beta\alpha}; \quad\, \bar{D}_{\alpha}=\bar{\partial}_{\alpha}.
\eea
These derivatives evidently anticommute with all generators. The commutators of these derivatives between themselves are
\bea
\{D_{\alpha},D_{\beta}\}=\{\bar{D}_{\alpha},\bar{D}_{\beta}\}=0; \quad\, \{D_{\alpha},\bar{D}_{\beta}\}=i\partial_{\alpha\beta}\equiv P_{\alpha\beta}.
\eea
The use of the "mixed" mutually conjugated complex spinor coordinates $\theta_{\alpha},\bar{\theta}_{\alpha}$ considered earlier in \cite{earlierHD}, instead of the usual real coordinates $\theta_{\alpha}^{1,2}$, allows to use the machinery of the well-developed $d=4$ superfield SUSY whose algebra is very similar to the algebra we use here (we note that in the previous paper by some of us \cite{earlier} namely the real spinor coordinates have been used).

Some more useful relations for spinor derivatives are
\bea
\label{idder}
&& \{D_{\alpha},\bar{D}_{\beta}\}=i\partial_{\alpha\beta};\quad\, \{D_{\alpha},D_{\beta}\}=\{\bar{D}_{\alpha},\bar{D}_{\beta}\}=0, \quad\, D_{\alpha}D^2=\bar{D}_{\alpha}\bar{D}^2=0; \nonumber\\
&& D^{\alpha}D_{\beta}=\delta^{\alpha}_{\beta}D^2; \quad\, \bar{D}^{\alpha}\bar{D}_{\beta}=\delta^{\alpha}_{\beta}\bar{D}^2; \quad\, [D^{\alpha},\bar{D}^2]=i\partial^{\alpha\beta}\bar{D}_{\beta}; \quad\, [\bar{D}^{\alpha},D^2]=i\partial^{\alpha\beta}D_{\beta};\nonumber\\
&& \bar{D}^2D^2\bar{D}^2=\Box \bar{D}^2;\quad\, D^2\bar{D}^2D^2=\Box D^2.
\eea
Also, one can have the projector-like identities \cite{earlierHD} which have no analogues in the four-dimensional superfield theories:
\bea
\label{usefulid}
(\bar D^\alpha D_\alpha)^n&=&\Box^{\frac{n-1}{2}}\bar D^\alpha D_\alpha \ , \ n=2l-1,\\
(\bar D^\alpha D_\alpha)^n&=&-\Box^{\frac{n}{2}-1}D^\alpha\bar D^2 D_\alpha \ , \ n=2l,\nonumber\\
(D^\alpha \bar D^2D_\alpha)^n&=&(-1)^{n+1}\Box^{n-1}D^\alpha \bar D^2D_\alpha \ , \ n=1,2,3,\ldots \ , \nonumber
\eea
where $l=1,2,3,\ldots$.

Then we suppose that one set of the fermionic coordinates, say $\theta^{\alpha}$, satisfy not Grassmann-like, but Clifford-like anticommutation relations:
\bea
\label{deformation}
\{\theta^{\alpha},\theta^{\beta}\}=2\Sigma^{\alpha\beta},
\eea
with all other anticommutators of spinor coordinates continue to be zero. These relations can be realized only on the Euclidean superspace, where $\theta^\alpha$ and $\bar\theta^\alpha$ are independent variables ($(\theta^\alpha)^\dagger\neq\bar\theta^\alpha$) \cite{KPT,Sei}. From now on, we will consider the ${\cal N}=2$ Euclidean superspace in this work. Here, $\Sigma^{\alpha\beta}$ is a constant matrix (actually, it introduces a constant vector by the rule $\Sigma^m=\frac{1}{\sqrt{2}}(\gamma^m)_{\alpha\beta}\Sigma^{\alpha\beta}$ breaking thus a Lorentz symmetry). This anticommutation relation can be implemented through replacement of all simple products of the spinor coordinates by the Moyal ones. We choose the Moyal product to be
\bea
\label{moyal1}
f(\theta)*g(\theta)=f(\theta)\exp(-\overleftarrow\partial_{\alpha}\Sigma^{\alpha\beta}\overrightarrow\partial_{\beta})g(\theta).
\eea
It follows from (\ref{Q's}) that this product has an equivalent form in terms of the SUSY generators
\bea
\label{moyal2}
f(\theta)*g(\theta)=f(\theta)\exp(\overleftarrow Q_{\alpha}\Sigma^{\alpha\beta}\overrightarrow Q_{\beta})g(\theta).
\eea
It is clear that the Moyal commutator is $\{\theta_{\alpha},\theta_{\beta}\}_*=2\Sigma_{\alpha\beta}$ which is consistent with our suggestion. It is also clear that the Moyal products (\ref{moyal1}) and (\ref{moyal2}) represent themselves as finite series for any superfields. Moreover, for any two superfields $\Phi_1$ and $\Phi_2$ one has
\bea
\label{iden3}
\int d^7z \Phi_1*\Phi_2=\int d^7z \Phi_1\Phi_2,
\eea
so, under the sign of the integral, for any two superfields, the Moyal product is equivalent to the usual product. Also, since $\{Q_{\alpha},Q_{\beta}\}=0$, this product will be associative.
We also note that since the supercovariant derivatives developed by us involve explicitly only multiplication by $\bar{\theta}^{\alpha}$, none of the identities (\ref{idder},\ref{usefulid}) is affected by the Moyal product. One should observe also that only the supersymmetry transformations involving the generators $Q_{\alpha}$ are consistent with the Clifford-like anticommutation relations for spinor coordinates, therefore, we can treat the supersymmetry in our case to be partially broken, as in the four-dimensional case, see f.e. \cite{KPT,Sei}. So, we can speak about ${\cal N}=2/2$ supersymmetry. One can notice that since the Moyal product (\ref{moyal1}) does not affect the usual bosonic coordinates, there is no UV/IR mixing in this kind of theories.

A chiral superfield $\Phi$ is defined to satisfy the differential constraint $\bar D_\alpha\Phi=0$. The general solution of this equation is given by
\bea
\Phi(y,\theta)=\phi(y)+\theta^\alpha\psi_\alpha(y)-\theta^2F(y).
\eea
Since the spinorial generator $Q_\alpha$ commutes with all covariant derivatives $D_M=(D_\alpha,\bar D_\alpha,\partial_{\alpha\beta})$, then it follows from (\ref{moyal2}) that the Moyal product $\Phi_1*\Phi_2$ is again a chiral superfield.

For the product of the same fields, one will have
\bea
\label{iden1}
\Phi*\Phi&=&\Phi^2-\frac{1}{2}\Sigma^2(\partial^2\Phi)^2 ,\\
\label{iden2}
\Phi*\Phi*\Phi&=&\Phi^3-\frac{1}{2}\Sigma^2\Phi(\partial^2\Phi)^2-\frac{1}{2}\Sigma^2(\partial^2\Phi^2)(\partial^2\Phi) ,
\eea
where $\Sigma^2\equiv\Sigma^{\alpha\beta}\Sigma_{\alpha\beta}$.

The antichiral superfield $\bar\Phi$ can be defined in the usual way by $D_\alpha\bar\Phi=0$, and its general solution is given by
\bea
\bar\Phi(\bar y,\bar\theta)=\bar\phi(\bar y)+\bar\theta^\alpha\bar\psi_\alpha(\bar y)-\bar\theta^2\bar F(\bar y),
\eea
where one is using the antichiral coordinates $\bar z^M=(\bar y^{\alpha\beta},\theta^\alpha,\bar\theta^\beta)$, with $\bar y^{\alpha\beta}=y^{\alpha\beta}-\frac{i}{2}(\theta^\alpha\bar\theta^\beta+\theta^\beta\bar\theta^\alpha)$. Because of the Moyal commutator
\bea
[\bar y^{\alpha\beta},\bar y^{\gamma\lambda}]_*=\frac{1}{2}(\Sigma^{\alpha\gamma}\bar\theta^\beta\bar\theta^\lambda+\Sigma^{\alpha\lambda}\bar\theta^\beta\bar\theta^\gamma
+\Sigma^{\beta\gamma}\bar\theta^\alpha\bar\theta^\lambda+\Sigma^{\beta\lambda}\bar\theta^\alpha\bar\theta^\gamma) ,
\eea
 the functions of the antichiral coordinate $\bar y$ must be multiplied according to:
\bea
\label{moyal3}
\bar f(\bar y)*\bar g(\bar y)=\bar f(\bar y)\exp\big[\frac{1}{4}\overleftarrow{\bar\partial}_{\alpha\beta}(\Sigma^{\alpha\gamma}\bar\theta^\beta\bar\theta^\lambda+\Sigma^{\alpha\lambda}\bar\theta^\beta\bar\theta^\gamma
+\Sigma^{\beta\gamma}\bar\theta^\alpha\bar\theta^\lambda+\Sigma^{\beta\lambda}\bar\theta^\alpha\bar\theta^\gamma)\overrightarrow{\bar\partial}_{\gamma\lambda}\big]\bar g(\bar y),
\eea
where $\bar\partial_{\alpha\beta}\equiv\frac{\partial}{\partial\bar y^{\alpha\beta}}$. It follows trivially from (\ref{moyal3}) that for the product of the same fields, one will have $\bar\Phi^n_*=\bar\Phi^n$. Therefore, one do not have corrections, due to the noncommutativity, for the product of the same antichiral superfields.

Now, it is the time to develop consistent field theory models. In principle, it is natural to adopt the examples of the models developed in \cite{earlierHD}, that is, the models whose actions formally reproduce the structure of the well-known $d=4$ superfield theories. We start with the scalar field theory:
\bea
\label{classic}
S=\int d^7z \Phi\bar{\Phi}-\int d^5z\Big(\frac{m}{2}\Phi^2+\frac{\lambda}{3}\Phi_*^{3}+\frac{g}{4}\Phi_*^{4}\Big)-\int d^5\bar z\Big(\frac{\bar m}{2}\bar\Phi^2+\frac{\bar\lambda}{3}\bar\Phi^{3}+\frac{\bar g}{4}\bar\Phi^{4}\Big).
\eea
We note that within this theory one can introduce (anti)chiral fields satisfying the relations $D_{\alpha}\bar{\Phi}=0$ and $\bar{D}_{\alpha}\Phi=0$ (to do it, one can follow the line proposed in the papers \cite{Buch,Buch1}). Actually, the similarity of expression of the derivatives in our case and usual four-dimensional theory establishes a whole similarity of the field theory models. The only difference is the fact that there is only one spinor representation, and, so, some extra commutation relations like $\{D^{\alpha},\bar{D}_{\alpha}\}=0$. Effectively, one can proceed with applying the Moyal product (\ref{moyal1}) in the coupling terms of this action which will yield (after an integration by parts)
\bea
\label{vert1}
\int d^5z \Phi_*^{3}&=&\int d^5z \Phi\cdot\Phi_*^{2}=\int d^5z\big[\Phi^3-\frac{1}{2}\Sigma^2\Phi(\partial^2\Phi)^2\big],\\
\label{vert2}
\int d^5z \Phi_*^{4}&=&\int d^5z \Phi\cdot\Phi_*^{3}=\int d^5z\big[\Phi^4-\Sigma^2\Phi^2(\partial^2\Phi)^2\big],
\eea
where we used the identities (\ref{iden1}), (\ref{iden2}), and (\ref{iden3}). We can rewrite the expressions above in terms of component fields
\bea
\label{vert3}
\int d^5z(\Phi_*^{3}-\Phi^3)&=&-\frac{1}{2}\Sigma^2\int d^3xF^3,\\
\label{vert4}
\int d^5z(\Phi_*^{4}-\Phi^4)&=&-2\Sigma^2\int d^3x(\phi F+\psi^2)F^2.
\eea
These corrections are invariant under the transformations induced by $Q_\alpha$ on the component fields, but they are not under the ones induced by $\bar Q_\alpha$:
\bea
&&\delta_\epsilon\phi=-\epsilon^\alpha\psi_\alpha \ , \ \delta_\epsilon\psi_\alpha=\epsilon_\alpha F \ , \ \delta_\epsilon F=0;\\
&&\delta_{\bar\epsilon}\phi=0 \ , \ \delta_{\bar\epsilon}\psi_\alpha=-\bar\epsilon^\beta i\partial_{\alpha\beta}\phi \ , \ \delta_{\bar\epsilon}F=-\bar\epsilon^\beta i{\partial^\alpha}_\beta\psi_\alpha ,
\eea
respectively. Therefore, we notice that the ${\cal N}=2$ supersymmetry is broken to the ${\cal N}=2/2$ supersymmetry due to the deformation (\ref{deformation}).

It is not difficult to rewrite (\ref{vert3}) and (\ref{vert4}) in terms of superfields and covariant derivatives, then we easily get
\bea
\label{vert5}
\int d^5z(\Phi_*^{3}-\Phi^3)&=&-\frac{1}{2}\Sigma^2\int d^5z\Phi(D^2\Phi)^2,\\
\label{vert6}
\int d^5z(\Phi_*^{4}-\Phi^4)&=&-\Sigma^2\int d^5z\Phi^2(D^2\Phi)^2.
\eea
In this paper, we are interested in performing some simple quantum computations. However, we point out that the new vertices (\ref{vert5}) and (\ref{vert6}) are not the most convenient to deal with quantum calculations, due to the fact that we are not able to use a operator $\bar D^2$, which arises from the chiral functional derivative $\frac{\delta j(z)}{\delta j(z^{\prime})}=\bar D^2\delta^7(z-z^{\prime})$, to rewrite chiral integrals as integrals over full superspace. This drawback is due to the fact that the integrands of the new vertices (\ref{vert1}) and (\ref{vert2}) are not chiral.

In the literature, there are two approaches to overcome these problems. The first one, which was proposed in \cite{Spurion}, consists of introducing a spurion superfield $U\equiv\Sigma^2\theta^2\bar\theta^2$, such that we can rewrite (\ref{vert3}) and (\ref{vert4}) as integrals over full superspace
\bea
\int d^5z(\Phi_*^{3}-\Phi^3)&=&-\frac{1}{2}\int d^7zU(D^2\Phi)^3,\\
\int d^5z(\Phi_*^{4}-\Phi^4)&=&-2\int d^7zU\big[\Phi(D^2\Phi)+\frac{1}{2}(D^\alpha\Phi)(D_\alpha\Phi)\big](D^2\Phi)^2.
\eea
The second one, which was proposed in \cite{Quiral,Quiral1}, consists of introducing a chiral superfield $Q^2\Phi=-F(y)$, such that we can rewrite (\ref{vert3}) and (\ref{vert4}) as integrals over chiral superspace
\bea
\int d^5z(\Phi_*^{3}-\Phi^3)&=&-\frac{1}{2}\Sigma^2\int d^5z\Phi(Q^2\Phi)^2,\\
\int d^5z(\Phi_*^{4}-\Phi^4)&=&-\Sigma^2\int d^5z\Phi^2(Q^2\Phi)^2.
\eea
Of course, both approaches are equivalent, since both lead to the same results (\ref{vert3}) and (\ref{vert4}). In this paper we will make use only of the second one. Therefore, the classical action (\ref{classic}) can be rewritten as
\bea
\label{mainaction}
S=S|_{\Sigma=0}+\frac{1}{2}\Sigma^2\int d^5z\Big[\frac{\lambda}{3}\Phi(Q^2\Phi)^2+\frac{g}{2}\Phi^2(Q^2\Phi)^2\Big].
\eea
The classical action in this form allow us to derive the Feynman rules for the pertubative expansion of the effective action by the straightforward application of the usual supergraph method \cite{Superspace}. The rules for the propagators
\bea
\label{propagador 1}
\langle\overline\Phi_1\Phi_2\rangle&=&\frac{1}{k^2+\bar mm}\delta_{12},\\
\label{propagador 2}
\langle\Phi_1\Phi_2\rangle&=&-\frac{\bar mD_1^2}{k^2(k^2+\bar mm)}\delta_{12},\\
\label{propagador 3}
\langle\overline\Phi_1\overline\Phi_2\rangle&=&-\frac{m\overline D_1^2}{k^2(k^2+\bar mm)}\delta_{12},
\eea
and for the vertices $\Phi^3$ and $\Phi^4$ are the same as those for the undeformed model. However, the rules for the vertices $\Phi(Q^2\Phi)^2$ and $\Phi^2(Q^2\Phi)^2$ have the
additional feature that for $(2-n)$ of the chiral internal lines leaving a vertex there is a factor $Q^2$ acting on the corresponding propagator, where $n$ is the number of $Q^2\Phi$-external lines in the respective vertex.

Let us start with two examples of quantum calculations contributing to the effective action. In the first example, we consider the supergraph of Fig. 1.

\begin{figure}[!h]
\begin{center}
\includegraphics[angle=0,scale=0.80]{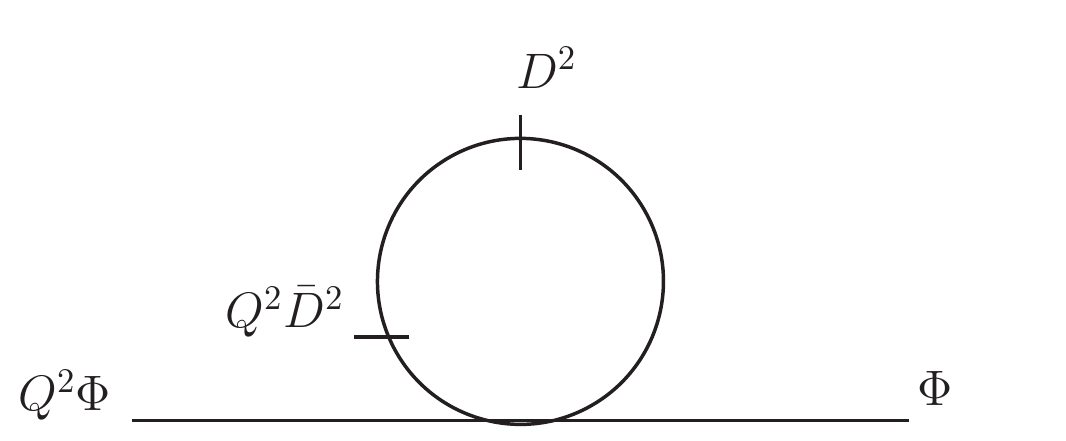}
\end{center}
\caption{Supergraph composed by one $\langle\Phi_1\Phi_2\rangle$-propagator and one $\Phi^2(Q^2\Phi)^2$-vertex.}
\end{figure}
It follows from the supergraph of Fig. 1
\bea
\Gamma_1[\Phi]=S_1\Big(-\frac{\Sigma^2g}{4}\Big)\int \frac{d^3p}{(2\pi)^3}d^4\theta_1\big[Q_1^2\Phi(-p,\theta_1)\big]\Phi(p,\theta_1)\int \frac{d^3k}{(2\pi)^3}\bigg[-\frac{\bar mQ_1^2\bar D_1^2D_1^2\delta_{12}|_{1=2}}{k^2(k^2+\bar mm)}\bigg],
\eea
where $S_1$ is the symmetry factor. By using the identity $Q_1^2\bar D_1^2D_1^2\delta_{12}|_{1=2}=-k^2\bar\theta_1^2$, we obtain
\bea
\Gamma_1[\Phi]=-\frac{1}{4}S_1\Sigma^2g\bar m\int \frac{d^3p}{(2\pi)^3}d^4\theta\bar\theta^2\big[Q^2\Phi(-p,\theta)\big]\Phi(p,\theta)\int \frac{d^3k}{(2\pi)^3}\frac{1}{(k^2+\bar mm)}.
\eea
After solving the integrals, we get
\bea
\label{cont1}
\Gamma_1[\Phi]=-\frac{1}{16\pi}S_1\Sigma^2g\bar m^{\frac{3}{2}}m^{\frac{1}{2}}\int \frac{d^3p}{(2\pi)^3}d^2\theta\big[Q^2\Phi(-p,\theta)\big]\Phi(p,\theta).
\eea
In our next example, we consider the supergraph of Fig. 2.

\begin{figure}[!h]
\begin{center}
\includegraphics[angle=0,scale=0.80]{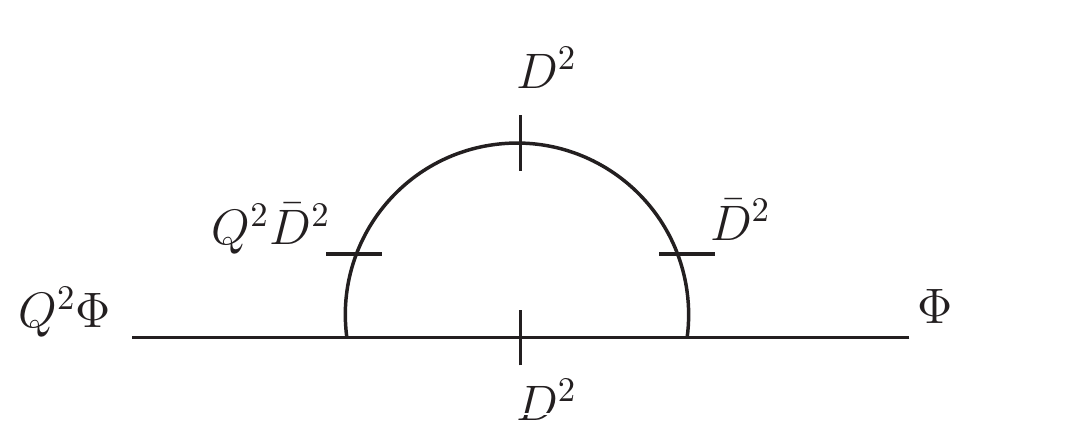}
\end{center}
\caption{Supergraph composed by two $\langle\Phi_1\Phi_2\rangle$-propagators, one $\Phi(Q^2\Phi)^2$-vertex, and one $\Phi^3$-vertex.}
\end{figure}

It follows from the supergraph of Fig. 2 that
\bea
\Gamma_2[\Phi]&=&S_2\Big(-\frac{\Sigma^2\lambda}{6}\Big)\Big(\frac{\lambda}{3}\Big)\int \frac{d^3p}{(2\pi)^3}d^4\theta_1d^4\theta_2\big[Q_1^2\Phi(-p,\theta_1)\big]\Phi(p,\theta_2)\nonumber\\
&\times&\int \frac{d^3k}{(2\pi)^3}\bigg[-\frac{\bar mD^2_2\delta_{21}}{k^2(k^2+\bar mm)}\bigg]\bigg[-\frac{\bar mQ_1^2\bar D_1^2D_1^2\bar D^2_1\delta_{12}}{(p+k)^2((p+k)^2+\bar mm)}\bigg],
\eea
where $S_2$ is the symmetry factor. Then using $\bar D_1^2D_1^2\bar D^2_1=-(p+k)^2\bar D^2_1$ and integrating the result by parts gives
\bea
\label{(33)}
\Gamma_2[\Phi]&=&\frac{1}{18}S_2\Sigma^2\lambda^2\bar m^2\int \frac{d^3p}{(2\pi)^3}d^4\theta_2\big[Q_2^2\Phi(-p,\theta_2)\big]\Phi(p,\theta_2)\nonumber\\
&\times&\int \frac{d^3k}{(2\pi)^3}\bigg[\frac{D^2_2Q^2_2\bar D^2_2\delta_{21}|_{2=1}}{k^2(k^2+\bar mm)}\bigg]\frac{1}{(p+k)^2+\bar mm}.
\eea
From $D_2^2Q_2^2\bar D_2^2\delta_{21}|_{2=1}=-k^2\bar\theta_2^2$ and (\ref{(33)}), we obtain
\bea
\Gamma_2[\Phi]=\frac{1}{18}S_2\Sigma^2\lambda^2\bar m^2\int \frac{d^3p}{(2\pi)^3}\frac{d^3k}{(2\pi)^3}\frac{1}{(k^2+\bar mm)[(p+k)^2+\bar mm]}\int d^2\theta\big[Q^2\Phi(-p,\theta)\big]\Phi(p,\theta).
\eea
Finally, integrating the result, we get
\bea
\label{cont2}
\Gamma_2[\Phi]=\frac{1}{72\pi}S_2\Sigma^2\lambda^2\bar m^2\int \frac{d^3p}{(2\pi)^3}\frac{1}{|p|}\arcsin{\Bigg[\frac{|p|}{(p^2+4\bar mm)^{\frac{1}{2}}}\Bigg]}\int d^2\theta\big[Q^2\Phi(-p,\theta)\big]\Phi(p,\theta).
\eea
We notice that the one-loop quantum corrections for the effective action (\ref{cont1}) and (\ref{cont2}) are UV-finite. This is a natural consequence of the low dimensionality of the spacetime, namely d=3. Moreover, we also notice that the corrections (\ref{cont1}) and (\ref{cont2}) are holomorphic. This result is consistent with the non-renormalization theorem for deformed theories \cite{Quiral,Quiral1}.

Now, let us study the low-energy effective action (LEEA). The LEEA is defined as the zero-order term in the covariant derivative expansion of the effective action of background superfields. Here, our goal is to calculate the one-loop correction to the LEEA within the model (\ref{mainaction}). In order to evaluate it we have to expand (\ref{mainaction}) around a background superfield $\Phi\rightarrow\Phi+\phi$ and $\bar\Phi\rightarrow\bar\Phi+\bar\phi$, and to keep the quadratic terms in quantum fluctuations $\phi$ and $\bar\phi$. Therefore, one can show that the one-loop contribution to $\Gamma[\Phi,\bar\Phi]$ is given by \cite{BuKu,Grisaru}:
\bea
\label{1loopEA}
\Gamma^{(1)}=-\frac{1}{2}\textrm{Tr}\ln\hat{\mathcal H}=-\frac{1}{2}\int d^7z\textrm{tr}\ln\hat{\mathcal H}\delta^7(z-z^{\prime})|_{z=z^\prime}.
\eea
where $\textrm{tr}$ is the matrix trace and $\hat{\mathcal H}$ is an operator that is obtained from the quadratic terms in quantum fluctuations in the classical action $S[\Phi+\phi,\bar\Phi+\bar\phi]$ that we will call of $S_2[\Phi,\bar\Phi;\phi,\bar\phi]$. Therefore, by using this prescription, we obtain from (\ref{mainaction}), after an integration by parts,
\bea
\label{quadratic}
S_2[\bar\Phi,\Phi;\bar\phi,\phi]&=&\int d^7z\bar\phi\phi-\frac{1}{2}\int d^5\bar z\bar\phi(\bar m+2\bar\lambda\bar\Phi+3\bar g\bar\Phi^2)\bar\phi-\frac{1}{2}\int d^5z\phi\Big\{m+2\lambda\Phi+3g\Phi^2\nonumber\\
&&-\frac{\Sigma^2g}{2}(Q^2\Phi)^2-\Sigma^2\big[\lambda(Q^2\Phi)+\frac{g}{2}(Q^\alpha\Phi)(Q_\alpha\Phi)+3g\Phi(Q^2\Phi)\big]Q^2\Big\}\phi.
\eea
It is convenient to write the antichiral and chiral superfields in terms of unconstrained superfields, namely $\Phi=\bar D^2\Psi$, $\bar\Phi=D^2\bar\Psi$, and $\phi=\bar D^2\psi$, $\bar\phi=D^2\bar\psi$ \cite{Grisaru}. Putting them in (\ref{quadratic}), it follows that (\ref{quadratic}) is invariant under the following Abelian gauge transformation $\delta\Psi=\bar D^{\dot{\alpha}}\bar\omega_{\dot{\alpha}}$, and $\delta\bar\Psi=D^{\alpha}\omega_{\alpha}$. Therefore, to fix the gauge, we add the functional \cite{Superspace}
\bea
\label{ghostfixing}
S_{GF}[\psi,\bar\psi]=\int d^7z \bar\psi(\frac{1}{2}\{\bar D^2,D^2\}-D^\alpha\bar D^2D_\alpha)\psi \ .
\eea
Since the theory is Abelian, the ghosts associated with this gauge fixing decouple. With all this information, we can sum up (\ref{ghostfixing}) and (\ref{quadratic}) to get
\bea
S_2[\Phi,\bar\Phi;\psi,\bar\psi]+S_{GF}[\psi,\bar\psi]=\frac{1}{2}\int d^7z
\left(\begin{array}{cc}
\psi & \bar\psi
\end{array}\right)
\hat{\mathcal H}
\left(\begin{array}{c}
\psi \\
\bar\psi
\end{array}\right) \ ,
\eea
where,
\bea
\hat{\mathcal H}=
\left(\begin{array}{cc}
-(\mu-\Sigma^2\mu_K-\Sigma^2\mu_WQ^2)\bar D^2 & \Box \\
\Box & -\bar\mu D^2
\end{array}\right) \ ,
\eea
and
\bea
\label{mu1andmu2}
\mu=m+2\lambda\Phi+3g\Phi^2 \ , \ \mu_K=\frac{g}{2}(Q^2\Phi)^2,\\
\label{mu3}
\mu_W=(\lambda+3g\Phi)(Q^2\Phi)+\frac{g}{2}(Q^\alpha\Phi)(Q_\alpha\Phi).
\eea
It is convenient to split $\hat{\mathcal H}$ into two parts, namely $\hat{\mathcal H}\equiv\hat{\mathcal H}_K+\hat{\mathcal H}_W$. Therefore, we can rewrite (\ref{1loopEA}) as
\bea
\label{EA}
\Gamma^{(1)}=-\frac{1}{2}\textrm{Tr}\ln(\hat{\mathcal H}_K+\hat{\mathcal H}_W)=-\frac{1}{2}\textrm{Tr}\ln\hat{\mathcal H}_K-\frac{1}{2}\textrm{Tr}\ln(\hat1+\hat{\mathcal H}_K^{-1}\hat{\mathcal H}_W)\equiv\Gamma_K^{(1)}+\Gamma_W^{(1)},
\eea
where
\bea
&&\hat{\mathcal H}_K=
\left(\begin{array}{cc}
-(\mu-\Sigma^2\mu_K)\bar D^2 & \Box \\
\Box & -\bar\mu D^2
\end{array}\right) \ , \
\hat{\mathcal H}_W=
\left(\begin{array}{cc}
\Sigma^2\mu_WQ^2\bar D^2 & 0 \\
0 & 0
\end{array}\right) \ ,\\
&&\hat{\mathcal H}^{-1}_K=\frac{1}{\Box}
\left(\begin{array}{cc}
\displaystyle\bar\mu\frac{D^2}{\Box-(\mu-\Sigma^2\mu_K)\bar\mu} &  \displaystyle1+\frac{\bar\mu(\mu-\Sigma^2\mu_K)D^2\bar D^2}{\Box\big[\Box-\bar\mu(\mu-\Sigma^2\mu_K)\big]}\\
\displaystyle1+\frac{(\mu-\Sigma^2\mu_K)\bar\mu\bar D^2D^2}{\Box\big[\Box-(\mu-\Sigma^2\mu_K)\bar\mu\big]} & \displaystyle(\mu-\Sigma^2\mu_K)\frac{\bar D^2}{\Box-\bar\mu(\mu-\Sigma^2\mu_K)}
\end{array}\right) \ .
\eea
The first term in (\ref{EA}) can be calculated following the same steps as in \cite{Grisaru}, then we can simply repeat all this procedure. Therefore, after the calculation of the matrix and functional trace, we get
\bea
\Gamma_K^{(1)}=-\frac{1}{2}\textrm{Tr}\ln\hat{\mathcal H}_K=\frac{1}{2}\int d^7z\int \frac{d^3k}{(2\pi)^3}\frac{1}{k^2}\ln\Big[1+\frac{\bar\mu(\mu-\Sigma^2\mu_K)}{k^2}\Big] \ .
\eea
By integrating and substituting the expressions (\ref{mu1andmu2}), we get
\bea
\label{Kpotential}
\Gamma_K^{(1)}=\frac{1}{4\pi}\int d^7z\Big[\big(\bar m+2\bar\lambda\bar\Phi+3\bar g\bar\Phi^2\big)\big(m+2\lambda\Phi+3g\Phi^2-\frac{1}{2}\Sigma^2g(Q^2\Phi)^2\big)\Big]^{\frac{1}{2}}.
\eea
This result corresponds to the one-loop K\"{a}hler effective potential, it is UV-finite, and we notice that the functional structure of (\ref{Kpotential}) does not
involve any logarithmlike dependence, which is usually found in four-dimensional theories. One can point out that the one-loop finiteness is a characteristic feature of the three-dimensional (in particular, supersymmetric) theories, see f.e. \cite{ourEP}. Besides, in the limit $\Sigma\rightarrow0$, we recover the one-loop K\"{a}hler effective potential for the commutative theory studied in \cite{BMS}. It is worth to point out that if we had "turned off" the quartic interaction ($g=0$), we would have got the undeformed result in (\ref{Kpotential}). In other words, we would have got the one-loop K\"{a}hler effective potential with unbroken ${\mathcal N}=2$ supersymmetry.

Let us move on and calculate the second term in (\ref{EA}), then we have
\bea
\Gamma_W^{(1)}&=&-\frac{1}{2}\textrm{Tr}\ln(\hat1+\hat{\mathcal H}_K^{-1}\hat{\mathcal H}_W)\nonumber\\
&=&-\frac{1}{2}\textrm{Tr}\ln\Bigg[\left(\begin{array}{cc}
1 & 0 \\
0 & 1
\end{array}\right)+\left(\begin{array}{cc}
\displaystyle\frac{\Sigma^2\bar\mu\mu_W}{\Box-(\mu-\Sigma^2\mu_K)\bar\mu}Q^2\frac{D^2\bar D^2}{\Box} & 0 \\
\displaystyle\frac{\Sigma^2\mu_W}{\Box-(\mu-\Sigma^2\mu_K)\bar\mu}Q^2\bar D^2 & 0
\end{array}\right)\Bigg].
\eea
By calculating the matrix trace and writing the integral over whole superspace as $\int d^7z=\int d^5z\bar D^2$, we get
\bea
\label{Wpotentialex}
\Gamma_W^{(1)}&=&-\frac{1}{2}\textrm{Tr}\Bigg[\ln\bigg(1+\frac{\Sigma^2\bar\mu\mu_W}{\Box-(\mu-\Sigma^2\mu_K)\bar\mu}Q^2\bigg)\frac{D^2\bar D^2}{\Box}\Bigg]\nonumber\\
&=&-\frac{1}{2}\int d^5z\int\frac{d^3k}{(2\pi)^3}\ln\bigg(1-\frac{\Sigma^2\bar\mu\mu_W}{k^2+(\mu-\Sigma^2\mu_K)\bar\mu}Q^2\bigg)\bar D^2\delta^4(\theta-\theta^{\prime})|_{\theta=\theta^{\prime}}.
\eea
This is the exact result. However, the integral in (\ref{Wpotentialex}) is very complicated due to the fact that we are considering $Q_\alpha\Phi\ne0$. In order to perform the integral above, we will consider the approximation in which $\Sigma$ is very small. This approximation is reasonable, since we are calculating the LEEA and at low-energies is expected that $\Sigma$ be small. Therefore, we can rewrite (\ref{Wpotentialex}) as
\bea
\label{(50)}
\Gamma_W^{(1)}&=&\frac{1}{2}\Sigma^2\int d^5z\int\frac{d^3k}{(2\pi)^3}\frac{\bar\mu\mu_W}{k^2+\bar\mu\mu}Q^2\bar D^2\delta^4(\theta-\theta^{\prime})|_{\theta=\theta^{\prime}}+{\mathcal O}(\Sigma^4)\nonumber\\
&=&-\frac{1}{8\pi}\Sigma^2\bar\mu^{\frac{3}{2}}\int d^5z\mu^{\frac{1}{2}}\mu_W+{\mathcal O}(\Sigma^4).
\eea
Finally, turning off the antichiral superfield ($\bar\Phi=0\Rightarrow\bar\mu=\bar m$), we obtain
\bea
\label{Wpotentialap}
\Gamma_W^{(1)}=-\frac{1}{8\pi}\Sigma^2\bar m^{\frac{3}{2}}\int d^5z\big(m+2\lambda\Phi+3g\Phi^2\big)^{\frac{1}{2}}\big[(\lambda+3g\Phi)(Q^2\Phi)+\frac{g}{2}(Q^\alpha\Phi)(Q_\alpha\Phi)\big]+{\mathcal O}(\Sigma^4).
\eea
This result corresponds to the one-loop chiral effective potential. We notice that in the limit $\Sigma\rightarrow0$, we get from (\ref{Wpotentialex}) and (\ref{Wpotentialap}), $\Gamma_W^{(1)}=0$,  which is consistent with the vanishing of the one-loop chiral effective potential for the commutative theory, as argued in \cite{BMS}. From the result (\ref{Wpotentialex}), we can infer that the higher-order terms in (\ref{Wpotentialap}) are UV-finite. The argument is similar
to that used in \cite{TO} and goes as follows: if we expand the logarithm in (\ref{Wpotentialex}), only one of the factors $Q^2$ will act on $\bar D^2\delta^4(\theta-\theta^{\prime})|_{\theta=\theta^{\prime}}$ due to the fact that $Q^3=0$, so that $Q^2\bar D^2\delta^4(\theta-\theta^{\prime})|_{\theta=\theta^{\prime}}=1$. The other factors will act on $G(k^2)=\frac{1}{k^2+(\mu-\Sigma^2\mu_K)\bar\mu}$, so that every time that $Q^2$ acts on $G(k^2)$ the UV convergence will not be worsened, on the contrary, $Q_\alpha G(k^2)\sim\frac{1}{k^4}$. Since (\ref{(50)}) is UV-finite and every time that $Q^2$ acts on $G(k^2)$ the UV-convergence will be improved, it follows that the higher-order terms in (\ref{Wpotentialap}) will be UV finite.

We developed a new formulation of the three-dimensional noncommutative superspace which turns out to be more convenient for the superfield calculation than those ones described in \cite{earlier}. Within this description, one can straightforwardly use all methods and approaches which have been well developed for the four-dimensional noncommutative superfield theories. In particular, just as it takes place in the four-dimensional field theories, one should introduce a partial breaking of the ${\cal N}=2$ supersymmetry to implement the fermionic noncommutativity. We have constructed the chiral superfield model within this approach and explicitly calculated the two-point function and the one-loop low-energy effective action for this theory.

It is interesting to note that in this theory, from a formal viewpoint one meets the Lorentz symmetry breaking introduced through a constant matrix $\Sigma^{\alpha\beta}$ which is equivalent to a constant vector $\Sigma^m=\frac{1}{\sqrt{2}}(\gamma^m)_{\alpha\beta}\Sigma^{\alpha\beta}$. However, within this paper, all quantum corrections we found depend on the noncommutativity matrix only implicitly, through a scalar factor $\Sigma^2=\Sigma_{\alpha\beta}\Sigma^{\alpha\beta}$. This resembles the fact that within other Lorentz-breaking extension of the superfield approach, that is, the so-called aether superspace \cite{aether,aether1}, the Lorentz-breaking parameters enter the final result also through a scalar factor representing itself as a determinant of some matrix depending on the Lorentz-breaking parameter. This situation is typical for the contributions to the effective potential which do not involve derivatives of background superfields, as well as for some other low-order contributions which were considered in our paper. Nevertheless, in principle the contributions in which the noncommutativity matrix itself, and not only the scalars constructed on its base, enters, are possible, for example, the terms like $\Sigma^{\alpha\beta}D_{\alpha}\Phi\bar{D}_{\beta}\bar{\Phi}$, so, in general, the terms depending on the Lorentz-breaking parameter in an explicit form are not excluded.

The natural continuation of this study would consist in introducing the supergauge symmetry and models for supergauge field, and in studying of higher loop corrections, within this methodology. We are planning to do this in a forthcoming paper.

{\bf Acknowledgements.}  This work was partially supported by Conselho
Nacional de Desenvolvimento Cient\'{\i}fico e Tecnol\'{o}gico (CNPq). The work by A. Yu. P. has been supported by the CNPq project No. 303783/2015-0.

\end{document}